**Einleitung**

Hilary Putnams Biographie und philosophische Entwicklung spiegeln die Geschichte der angelsächsischen Philosophie in den letzten 40 Jahren. Beinahe ebenso lange hat Putnam diese Geschichte wesentlich beeinflußt und so kann John Passmore über Putnam schreiben: «Er ist die Geschichte der gegenwärtigen Philosophie im Umriß»[1].

In der vorliegenden Einleitung soll vor allem der Kontext dargestellt werden, in dem Putnam steht und aus dem heraus verständlich wird, was er philosophisch zu sagen hat. Dieser Kontext ist sicherlich ein Grund dafür, daß Putnam hierzulande noch relativ wenig bekannt ist, während er in den USA häufig für den bedeutendsten aktiven Philosophen gehalten wird. Im Rahmen einer Skizze von Putnams philosophischer Entwicklung soll zudem eine vorläufige philosophiehistorische Einordnung versucht werden, auch wenn hier nicht der Ort für eine umfassende Kritik oder Darstellung sein kann: Die Einleitung muß auf recht elementarem Niveau bleiben und kann eine Lektüre der Texte natürlich nicht ersetzen. Da Putnams Werk sicherlich Teil einer Annäherung von ‹analytischer› und ‹kontinentaler› Philosophie ist, soll bei der Einführung in die hier übersetzten Texte schließlich deutlich werden, was Putnam nicht analytisch orientierten Lesern zu bieten hat. (Diesen sei empfohlen, mit der Lektüre bei Text 8 zu beginnen, sofern keine speziellen thematischen Präferenzen vorhanden sind.)

Zunächst zur Vorgeschichte der hier übersetzten Texte: Die Philosophie im Amerika der Nachkriegszeit wurde nicht unwesentlich von Denkern geprägt, die vor dem Nationalsozialismus geflüchtet und nicht wieder in ihre Heimatländer zurückgekehrt waren. Zu diesen gehörte auch Hans Reichenbach, der in Berlin eine dem Wiener Kreis nahestehende Gruppe geleitet hatte und 1938 nach einigen Jahren in Istanbul an die *University of California, Los Angeles* (UCLA) wechselte. Reichenbach befaßte sich mit Grundlagenproblemen der Naturwissenschaften, insbesondere mit der philosophischen Bedeutung der neuen Quantenphysik. Putnam wurde sein Schüler und verfaßte 1951 eine Dissertation über *The Meaning od the Concept of Probability in Application to Finite Sequences* (1990, I-13[2]). Damit hatte er sich in die beiden zentralen Themen der analytischen Philosophie jener Zeit eingearbeitet: Mathematische Logik und das Problem der Bestätigung wissenschaftlicher Theorien, welches eng mit dem Begriff der Wahrscheinlichkeit verbunden ist.

Schon 1953 konnte Putnam eine Stelle als *Assistant Professor* an der renommierten Universität Princeton antreten. Dort traf er einen weiteren wichtigen Mentor: Rudolf Carnap. Dieser Philosoph lehrte seit 1936 in den USA und galt inzwischen als einer der führenden Köpfe im Lande. In den folgenden beiden Jahren führte Putnam zahlreiche Gespräche mit Carnap in dessen Haus. Neben den Klassikern des Pragmatismus (Peirce, Dewey, James) hat Putnam später immer wieder auf die Bedeutung seiner beiden Lehrer hingewiesen. In Princeton begann er seine Publikationstätigkeit (wie in der Bibliographie in diesem Band nachzulesen) mit Aufsätzen zu sprachphilosophischen Themen, zum Problem der Analytizität und zur Philosophie der Mathematik; außerdem griff er mit der Frage, ob die Quantenmechanik eine neue, nicht-klassische Logik erfordert (eine ‹Quantenlogik›), ein Thema Reichenbachs auf.

Mit dieser Fragestellung reihte sich Putnam bereits in eine Entwicklung gegen den beherrschenden Logischen Empirismus ein; denn dem Positivismus zufolge war Logik ja analytisch, und also konnte kein Ergebnis irgendeiner empirischen Wissenschaft für Logik relevant sein. Nach dem Logischen Empirismus war alle Erkenntnis auf Empfindungen der Sinne (‹Sinnesdaten›) zurückzuführen, und was darauf nicht zurückzuführen war, mußte entweder analytisch sein (Mathematik und Logik) oder als sinnlos gelten (‹Metaphysik›). Putnam hingegen meint inzwischen, es gebe auch in der Mathematik empirische Elemente. Damit beginnt er, einen ‹realistischen› Standpunkt einzunehmen, was hier zunächst einmal bedeutet, die Wissenschaften nicht als bloße Instrumente für erfolgreiche Voraussagen anzusehen.

Von Princeton ging Putnam 1961 an das *Massachusetts Institute of Technology* (MIT) in Boston. Sein wesentlicher praktischer Beitrag bestand in jener Zeit darin, den philosophischen Fachbereich am MIT um einen Graduiertenstudiengang zu erweitern, aus dem im Laufe der Jahre einige der bedeutendsten amerikanischen Philosophen hervorgegangen sind. Das MIT hat sich seitdem zu einem der führenden philosophischen Institute in den USA entwickelt. Putnams Interesse galt dort neben der Mathematik der allgemeinen Wissenschaftstheorie und vor allem der Philosophie des Geistes. Auch letztere stellte ja eine Schwachstelle im positivistischen System dar, insofern man sich gezwungen sah, Aussagen über den Verstand, die Psyche, den Geist (engl. *the mind*) anderer Menschen auf beobachtbares Verhalten zurückzuführen. Nur so schienen sie auf ‹öffentlich› zugängliche Anhaltspunkte reduzierbar und damit verifizierbar.

1966 verlegte Putnam seinen Schreibtisch einige Meilen flußaufwärts: Er erhielt einen Ruf auf den Lehrstuhl des *Walter Beverly Pearson Professor of Modern Mathematics and Mathematical Logic* an der Harvard Universität, wo er bis heute lehrt. In Harvard hat vor allem der um etwa 20

Jahre älterer Kollege Willard van Orman Quine wesentlichen Einfluß auf die Entwicklung der angelsächsischen Philosophie ausgeübt. Auch seine Philosophie kann als Reaktion auf den ‹importierten› Positivismus gedeutet werden, allerdings weniger als Abwendung, sondern eher als eine Fortentwicklung (bis zum bitteren Ende). In Harvard hat Putnam einen stetigen Strom von Aufsätzen publiziert, die gemeinsam einen fortwährenden Angriff auf den Positivismus bilden. Erst 1975 allerdings wurden die ersten beiden Bände *Philosophical Papers* zusammengestellt (1975, I-3 & 4).

Putnams selbst sieht sich als einen der ‹Überwinder› des Positivismus. Worin diese Überwindung besteht, sollte aus den hier übersetzten Texten deutlich werden. Es zeigt sich aber bereits an dieser Stelle ein wesentlicher Grund für die relative Unbekanntheit Putnams in Deutschland: Er hat sich mit der Kritik einer Philosophie befaßt, die hierzulande nie so recht Fuß fassen konnte (wohlgemerkt des Positivismus, nicht der ‹analytischen Philosophie› allgemein). Bestünde darin seine einzige Leistung, dann könnte man seine Schriften beruhigt unübersetzt lassen. Es werden in dieser Kritik aber beileibe nicht nur Dinge entwickelt, die für (Ex-)Positivisten von Interesse sind. Es wird eine ganze Bedeutungstheorie entfaltet, und die Kritik ist wesentlich konstruktiv, sie baut eine alternative philosophische Richtung auf: Realismus. Die Verteidigung und Weiterentwicklung dieser Alternative nimmt den wesentlichen Teil des vorliegenden Bandes ein (und in dieser Ausarbeitung wird eine Reihe von ‹großen Fragen› behandelt). Eine weitere, eher praktische Schwierigkeit, ist das Fehlen eines ‹Hauptwerks› oder überhaupt einer Darstellung in Buchlänge. Das gesamte Schaffen besteht aus kürzeren Texten, auch wenn sich diese gelegentlich zu einem kohärenten Buch fügen.[3] Putnam übt seinen Einfluß vor allem durch Aufsatzpublikationen aus.

Es mußte also eine Auswahl getroffen werden, die einen Zugang gestattet und nichts Wesentliches ausläßt, was auch machbar erschien. In einem Band von zehn Texten (d. h. weniger als fünf Prozent der Veröffentlichungen) war allerdings eine gewisse thematische Einschränkung nicht zu vermeiden, und hier soll zumindest kurz erwähnt werden, welche Themen im vorliegenden Band nicht repräsentiert sind:

Da wären zunächst die zum Großteil in Band I der *Philosophical Papers* (1975, I-3) abgedruckten Arbeiten zur Philosophie der Mathematik, zur Quantenlogik und Wissenschaftstheorie im engeren Sinn.[4] Hiervon wäre die Quantenlogik (nicht Quantorenlogik!) als Vorschlag zur Behandlung der Quantenphysik noch am ehesten von Interesse; sie eignet sich aber wenig zur Einführung in Putnams Denken auch wenn sich in ihr, wie oben angedeutet, bereits ein realistischer Zug findet.[5] Gedanken zur Deutung der Quantenphysik allgemein ziehen sich durch das gesamte Werk Putnams und sind auch hier in Text 9, ‹Realismus mit menschlichem Antlitz›, Abschnitt ‹Realismus›, sowie in Text 10, ‹Irrealismus und Dekonstruktion›, Abschnitt ‹Die Bedeutung der begrifflichen Relativität›, zu finden. Die Schriften aus Band I der *Philosophical Papers* seien aber denjenigen ans Herz gelegt, die sich für die Rolle von Konvention und Logik in der Wissenschaft und umgekehrt für die Rolle von Konvention und Wissenschaft in der Logik interessieren.

Bis heute spielt die Philosophie des Geistes eine Rolle in Putnams Denken. Er hatte in den 60er Jahren an Gedanken von Alan Turing angeknüpft und vorgeschlagen, das Gehirn als ‹Turingmaschine› zu betrachten, als einen Gegenstand, der weder durch physikalische noch durch ‹seelische› Zustandsbeschreibungen noch durch Verhaltensbeschreibungen richtig zu erfassen ist (Materialismus, Dualismus bzw. Behaviourismus). Es kann nur abstrakt durch seine *funktionalen* Zustände charakterisiert werden. Ganz verschieden gebaute Apparate können funktional identisch sein (die gleiche ‹Turingmaschine› sein). So könnte auch ein Computer mit einem Gehirn funktional identisch sein, selbst wenn ihre ‹Hardware› ganz verschieden ist. Es ergibt sich dann die Frage, ob funktionale Zustände mit physikalischen Zuständen identifiziert werden können — eine Frage, die Putnam zunächst bejaht und später verneint hat. Thesen wie diese haben sich unter dem Namen (abstrakter) ‹Funktionalismus› zur zeitgenössischen Orthodoxie entwickelt und wurden auch von ‹materialistischen› Philosophen des Geistes in Beschlag genommen (insbesondere von Jerry Fodor, einem Schüler Putnams). Putnam selbst lehnt den Funktionalismus inzwischen ab (1988, I-11). Die Texte zu diesem Themenkreis wurden für einen in Vorbereitung befindlichen Band mit Putnams Philosophie des Geistes reserviert. Für die Zwischenzeit sei auf bereits vorhandene Übersetzungen verwiesen (III-2, 4 und 9).

Alle Schriften zu diesen Themen und zu einer ganzen Reihe von weiteren Fragen finden sich in der Bibliographie.

## 2

Wie im Inhaltsverzeichnis angedeutet, lassen sich die Texte dieses Bandes grob in drei Teile gliedern, wobei die inhaltliche mit der chronologischen Ordnung zusammenfällt. Der erste Teil befaßt sich mit der Kritik des Positivismus und der Kritik anderer Reaktionen auf den Positivismus, wie sie etwa mit den Namen Thomas Kuhn und Paul Feyerabend verbunden sind. Es handelt sich um Schriften aus den 70er Jahren, vor Putnams ‹Wende› vom Realismus zum ‹internen› Realismus, die sich allerdings in Text 3, «Was ist ‹Realismus›?» bereits andeutet.

Eine Schwierigkeit beim Zugang zu Putnams Texten von einer ‹kontinentaleuropäischen› Perspektive bereitet die ungewohnte Methodik. Es soll etwa um erkenntnistheoretische Problemen gehen, die Rede ist aber nur von ‹Bedeutung› oder ‹Referenz›. Das hängt zunächst mit einer Sichtweise zusammen, die in der angelsächsischen Philosophie seit langem so gängig ist, daß es im allgemeinen müßig erscheint, sie zu thematisieren: Philosophische Auffassungen werden als Theorien der Bedeutung charakterisiert. Wenn wir also zum Beispiel annehmen, die erkenntnistheoretische Position des Positivismus laute, daß alle Erkenntnis nur auf sinnlicher Wahrnehmung beruhen müsse, dann wird das auf ‹Angelsächsisch› so formuliert, daß ein Satz über einen physikalischen Gegenstand auf einen Satz über Sinneseindrücke *reduziert* werden muß. «Dort ist ein Apfel» *bedeutet* dann eigentlich «Wenn Du dort hinsiehst und die Bedingungen normal sind, dann wirst Du den-und-den Sinneseindruck haben» (oder etwas ähnliches; über die Details kann man sich danach trefflich streiten). Ein klassischer *Idealist* wird also unsere gewöhnlichen Sätze irgendwie auf Sätze über Ideen im Geist reduzieren wollen — dem steht in der englischen Terminologie üblicherweise der *Realist* gegenüber, der meint, die Welt besteht unabhängig von einem wahrnehmenden Geist und der unsere Sätze (über Äpfel und dergleichen) ‹für bare Münze› nimmt. Der Realist will ohne Reduktion auskommen. Wer also die ‹verifikationistische Theorie der Bedeutung› des Positivismus kritisiert wie Putnam in den Texten 1 bis 3, versucht damit, der ganzen Lehre den Boden zu entziehen. (Das heißt jedoch nicht, daß sich Realismus in einer Theorie der Bedeutung erschöpfen würde.)

Als weitere Schwierigkeit erweist sich, daß klassisch erkenntnistheoretische Probleme häufig über eine Diskussion von (Natur-) Wissenschaft behandelt werden. Es handelt sich dann nicht im Wissenschaftstheorie (‹philosophy of science›) im eigentlichen Sinne, sondern um eine Betrachtung von Wissen (Erkenntnis), wo es systematisch gewonnen wird. Diese als ‹Epistemologie› bezeichnete Fragestellung setzt lediglich einen anderen Akzent als Erkenntnistheorie, indem sie sich weniger mit dem Problem der Wahrnehmung befaßt.

Nach einigen früheren Anläufen[6] gelingt Putnam in ‹Erklärung und Referenz› (1973), Text 1, eine handfeste Kritik und alternative These. Er stellt realistische und ‹idealistische› (verifikationistische) Theorien der Bedeutung gegenüber und vergleicht ihre Antworten angesichts eines neuen Problems. Anhand verschiedener Beispiele aus der Geschichte der Wissenschaften zeigt er eine Schwierigkeit für solche (‹idealistischen›) Bedeutungstheorien auf, bei denen die Bedeutung unserer Ausdrücke davon abhängt, was wir wissen können. Um sein Lieblingsbeispiel anzuführen: Wenn die Bedeutung des Ausdruckes ‹Elektron› in einer Liste von Merkmalen bestünde, dann müßte diese Liste von uns anhand des vorhandenen Wissens über diese Teilchen erstellt worden sein. Nach der üblicherweise Russell (etwa 1912) und Frege (1892) zugeschriebenen Theorie der Bedeutung würde diese Liste den *Sinn* von ‹Elektron› darstellen und bestimmen, worauf der Ausdruck zutrifft, worauf er ‹referiert› — nämlich auf die Gegenstände, welche die dort aufgelisteten Bedingungen erfüllen. Das führt aber zu einer Schwierigkeit: Wenn Niels Bohr nun 1911 eine bestimmte Theorie jener Teilchen hatte (eine bestimmte Liste) und sich später herausstellt, daß diese Vorstellung fehlerhaft war (was ja wahrscheinlich ist), dann hat Bohr mit ‹Elektron› nicht referiert, denn kein Teilchen erfüllt die Bedingungen auf seiner Liste! Er hatte gar keine Theorie *von Elektronen*! Jede Kontinuität von Wissenschaft oder irgendeiner Erkenntnis scheint unmöglich geworden zu sein. Das ist eine schwere Hypothek für die positivistische Wissenschaftstheorie: Es scheint, sie könne den Erfolg der Wissenschaften nicht erklären — dieser wird zu einem Wunder. Eine weitere Schwierigkeit besteht darin, daß die Einträge auf der ‹Liste› jener Bedeutungstheorie zufolge Teil des Sinns von ‹Elektron› und also bloße Erläuterungen der Bedeutung von ‹Elektron› sind. «Elektronen haben negative Ladung» wäre dieser Ansicht zufolge keine empirische Erkenntnis, sondern ein analytisch wahrer Satz!

Diese Konsequenzen erscheinen unannehmbar, und Putnam versucht, eine ‹realistische› Theorie der Bedeutung zu formulieren, in der gewissermaßen die Elektronen selbst einen Einfluß auf die Bedeutung unseres Ausdrucks ausüben und Referenzkonstanz gewährleisten. «Anstatt Bedeutungen als Entitäten zu sehen, die Referenz bestimmen, sehen sie [die Realisten] Bedeutungen nun als weitgehend von Referenz, und Referenz als weitgehend von kausalen Verbindungen bestimmt.»[7] Er ist nicht bereit, Feyerabend zu folgen, der den Zusammenhang von Wissen verschiedener Zeiten und Kulturen aufgibt. Feyerabend bleibt bei einer extremen empiristischen Position und kommt im Endeffekt zu einem vollständigen Relativismus (er überwindet den Positivismus also nicht, sondern führt ihn intern fort — und *ad absurdum*, wie man sagen könnte). Putnam meint nun, zeigen zu können, daß die Bedeutungstheorie nicht von dieser mageren Kost leben muß und also nicht jede Theorie (oder ‹Kultur›) auf ihre eigenen Gegenstände referiert. Hilfreich in diesem Zusammenhang sind die Überlegungen von Saul Kripke (1972), der bei der Untersuchung von Eigennamen auf ähnliche Schwierigkeiten gestoßen ist und vorgeschlagen hat, die Referenz von Eigennamen müsse mit Hilfe einer kausalen Verbindung zwischen Träger und Namen erklärt werden. So hoffen die Realisten auf in der Welt schon vorhandene Arten und Gegenstände, die bei der Bestimmung von Referenz und Bedeutung helfen. Wie Putnam in Text 5 zusammenfaßt: «Multiple Sklerose, Gold, Pferde oder Elektrizität unterliegen *objektiven Gesetzen*, und was vernünftigerweise zu diesen Klassen zu zählen ist, wird davon abhängen, wie diese Gesetze sein werden.»

In ‹Sprache und Wirklichkeit› (1975), Text 2, formuliert Putnam zwei Maximen, denen eine Theorie der Bedeutung genügen soll. Diese sind insofern interessant, als sie weitgehend zum Gemeingut geworden sind und auch implizit schon in ‹Erklärung und Referenz› eine Rolle gespielt hatten. Es handelt sich um das Prinzip Vertrauensvorschuß und das Prinzip der sprachlichen Arbeitsteilung. Man sollte Sprechern einen Vertrauensvorschuß gewähren, das heißt annehmen, daß sie selbst dann auf Gegenstände referieren, wenn ihre Beschreibung derselben nicht ganz zutreffend ist. Wenn Bohr im Jahre 1911 also eine nicht ganz richtige Theorie von Elektronen hatte, dann sollte man dennoch annehmen, er referierte auf Elektronen — und nicht auf nichts. Wie kann dieser Vertrauensvorschuß aber gerechtfertigt werden? Die Grundlage hierfür ist, daß Bohrs Gebrauch des Wortes ‹Elektron› kausal mit Situationen verbunden ist, in denen das Wort eingeführt wurde. Diese Einführungssituationen stiften eine Verbindung zwischen Elektronen und ‹Elektron›, die zur Erklärung der Referenz genutzt werden kann.

Sprachliche Arbeitsteilung meint, daß die Referenz eines Ausdrucks nicht allein durch das Wissen eines einzelnen Sprechers bestimmt wird. Man gesteht bestimmten Experten aus der Sprachgemeinschaft eine Autorität zu und läßt sie entscheiden, was etwa als Elektron oder Goldstück zählen soll. Es sind dies nicht Linguisten, sondern Physiker oder Juweliere, Experten in der Sache. Wenn das richtig ist, kann die Bedeutung meiner Worte nicht allein durch meine Sinnesdaten bestimmt sein, sie kann keine Sache von Kriterien sein, über die ich allein die Autorität habe. Bedeutung kann nicht in meinem Kopf sein.

Im verbleibenden Teil des Textes versucht Putnam dann, die philosophische Fruchtbarkeit dieser Maximen zu zeigen, und baut sie zu einer Theorie der Referenz aus.[8]

Im Frühjahr 1976 hielt sich Putnam in England auf, um die *John Locke Lectures* in Oxford zu halten. Auch bei der Londoner *Aristotelian Society* hielt er bei dieser Gelegenheit einen Vortrag, der, wie dort üblich, schon vorher in deren *Proceedings* veröffentlicht worden war. Sein Titel lautete «Was ist ‹Realismus›?» Dieser dritte Text im vorliegenden Band stellte sich der Frage, was denn eigentlich aus den bedeutungstheoretischen Erwägungen folgen sollte. Hatten sie wirklich den Kern der Debatte ‹Realismus — Idealismus› getroffen? Und ihre Relevanz für den Begriff der Wahrheit? War dieser von Tarski ein-für-allemal definiert worden, oder hatte Tarskis Arbeit[9] das eigentliche Problem nur gestreift, wie Hartry Field (1972) meinte?

Zunächst wehrt Putnam sich gegen die Annahme, Realismus sei durch die Akzeptanz von Tarskischen Schemata ausreichend charakterisiert. Wenn eine Korrespondenztheorie der Wahrheit bloß Sätze wie «‹Schnee ist weiß› ist wahr dann und nur dann, wenn Schnee weiß ist» hervorbringt, kann sie allein Realismus nicht unterstützen. Zur Definition muß ein bestimmtes substantielles *Verständnis* der Referenz und Wahrheit kommen, in dem beide nicht von menschlichen Erkenntnismöglichkeiten abhängig gemacht werden. Das wäre etwa der Fall, wenn ein Idealist oder Instrumentalist (der Theorien nur als nützliche Instrumente für Voraussagen ansieht) meint, ein Satz *muß* wahr sein, wenn bestimmte uns zugängliche Kriterien erfüllt sind (‹die-und-die Sinneseindrücke sind vorhanden›, etwa) — seine Wahrheit besteht schließlich im Bestehen dieser Kriterien. Diesen Wahrheitsbegriff lehnt der Realist als bloßes empiristisches Surrogat ab. Seiner Ansicht nach geht Wahrheit über prinzipielle Erkennbarkeit hinaus.

Weiter muß ein Realist die oben angedeuteten Konsequenzen idealistisch-positivistischer Bedeutungstheorien — radikaler Referenzwechsel bei aufeinander folgenden Theorien, keine Konvergenz des wissenschaftlichen Wissens — als unannehmbar ansehen, was etwa Kuhn oder Feyerabend nicht ohne weiteres tun würden.[10]

## 3

Die Lehre des Realismus schien zu diesem Zeitpunkt ausreichend charakterisiert zu sein, und so wurde auch ihre dezidierte Kritik erst möglich. Die Textes dieses Bandes können als Einführung in die Debatte ‹Realistmus — Antirealismus› gelesen werden. Michael Dummett hatte bis Mitte der 70er Jahre eine Unterscheidung von Realismus und Antirealismus präzisiert, die auf semantischen Charakteristika beruhte.

Der Antirealist für einen bestimmten Bereich des Diskurses (etwa über physikalische Gegenstände) behauptet, daß die Sätze dieses Bereichs auf Sätze reduziert werden müssen, die verifizierbare Bedingungen ergeben (etwa das Vorliegen bestimmter Sinnesdaten), welche die Behauptung des Satzes rechtfertigen. Putnam wurde in diesen Jahren deutlich von Dummett beeinflußt und begann, seinen ursprünglichen Realismus zu modifizieren, was schließlich zu einer ‹Wende› führte. Seine Versuche in dieser Richtung werden hier nur in ihrer reiferen Form berücksichtigt.[11] Es sei noch betont, daß eine Reihe von Anhängern des frühen Putnam dieser Wende nicht gefolgt sind und ihn für einen Abtrünnigen halten — ganz wie es ja auch bei anderen Philosophen der Fall gewesen ist.

Wichtig ist hier zunächst das modelltheoretische Argument aus ‹Modelle und Wirklichkeit› (1980), Text 4. Putnam prüft die Frage, ob Realismus tatsächlich in der Lage ist, eine bestimmte Referenzrelation für einen Ausdruck als die richtige auszuzeichnen — und er kommt zu einem negativen Ergebnis. Wie kann man sicherstellen, daß ‹Gold› nur auf einen bestimmten Stoff in der Welt referiert, wie kann man zwischen dem Wort und dem Stoff jene scheinbar magische Verbindung der ‹Referenz› herstellen? Putnam bietet den ganzen Apparat der Modelltheorie auf, um zu zeigen, daß immer viele Interpretationen möglich sind. Ob dieser Apparat wesentlich zum Argument beiträgt, ist in der Literatur strittig und kann hier nicht diskutiert werden.[12] Die Schwierigkeit für den Realisten besteht darin, daß es mehrere ‹wahre› Theorien über die Welt geben kann und keine Entscheidung zwischen diesen möglich oder nötig ist.

Er zeigt, daß es zu jeder Theorie eine andere gibt, die gegenüber der ersten keinen epistemischen Unterschied macht (beide Theorien führen unter allen Umständen zu genau denselben Voraussagen), aber etwas anderes sagt. Man könnte die zweite Theorie auch bloß als eine andere ‹Interpretation› der ersten ansehen. Nun, wäre es nicht möglich, eine Theorie aufstellen, welche festlegt, welche Interpretation ‹gemeint› war?

Putnam behauptet, das ist nicht möglich, denn auch diese Theorie hätte wiederum mehrere ‹Interpretationen› und so fort, *ad infinitum*. Man würde so lediglich mehr Theorie produzieren, aber das Problem nicht lösen. Der Realist kann nur noch einen ‹metaphysischen Klebstoff› postulieren, der Referenz auf magische Weise festlegt.

Putnam allerdings will sich nicht einfach auf die Seite der Antirealisten schlagen, denn er hält Realismus *innerhalb* einer Theorie nach wie vor für angemessen. Innerhalb der Theorie sind Gebrauch und Referenz verknüpft; es ist festgelegt, was als Gegenstand gelten kann und was nicht, und also kann das modelltheoretische Argument nicht angewendet werden. Putnams neue Position heißt *interner Realismus*, und die verworfene Variante des Realismus wird als *metaphysischer Realismus* gebrandmarkt. In der Literatur wird der ‹neue› Putnam allerdings sehr häufig als Antirealist bezeichnet, denn er wendet sich ja gegen die ursprüngliche Orthodoxie des Realismus. Außerdem bestreitet er, daß eine epistemisch ideale Theorie noch sinnvoll als falsch bezeichnet werden kann — und das wird häufig als Kennzeichen des Antirealisten gesehen.

In ‹Referenz und Wahrheit› (Text 5), einem Artikel für eine italienische Enzyklopädie, hat Putnam die Entwicklung seines Denkens bis 1980 noch einmal zusammengefaßt. Er beschreibt sowohl die frühere Referenztheorie und ihre Bedeutung als auch die Wende und sein Verhältnis zu Tarski, Davidson und Dummett. Dieser Text ist trotz seiner thematischen Dichte sicherlich zugänglicher als die vorhergehenden vier und er bietet zudem eine konzise Darstellung von Tarskis Wahrheitstheorie. Er könnte als Einleitung zu den Texten 1 bis 4 gelesen werden und widerlegt zugleich die Auffassung, mit der Wende zum internen Realismus habe Putnam auch die Einsichten seiner früheren ‹kausalen› Referenztheorie aufgegeben.

Putnam distanziert sich hier von Dummett und betont, daß er Bedeutung nicht mit einer Liste von verifizierbaren Bedingungen identifiziert wissen will, welche die Behauptung eines Satzes rechtfertigen würden. Wahrheit würde dann zu Berechtigung, sie sollte aber *idealisierte* Berechtigung sein.[13] Bedeutung (und damit Wahrheit) kann nicht vollkommen unabhängig von menschlichen epistemischen Möglichkeiten sein, sollte aber nicht auf gegenwärtig für uns verfügbare Anhaltspunkte eingeschränkt werden. Ohnehin will Putnam nicht wieder von einem Reduktionismus eingefangen werden, wie er eigentlich für Non-Realisten typisch ist.

Der interne Realismus bedarf zu diesem Zeitpunkt noch einiger Präzisierungen, und zu diesem Zweck greift Putnam auf klassische Literatur der Erkenntnistheorie zurück (Locke, Berkeley, Kant). In ‹Wie man zugleich interner Realist und transzendentaler Realist sein kann› (1980, Text 6) versucht er, die Ähnlichkeit von internem Realismus und Kants transzendentalem Idealismus zu ergründen. Dazu formuliert er noch einmal die Thesen, die für metaphysischen Realismus charakteristisch sein sollen, um sich von ihm abzugrenzen: «Die Welt besteht aus einer feststehenden Gesamtheit geistunabhängiger Gegenstände. Es gibt genau eine wahre und vollständige Beschreibung davon, ‹wie die Welt ist›; Wahrheit beinhaltet eine Art von Entsprechungsbeziehung zwischen Wörtern oder Gedankenzeichen und äußeren Dingen sowie Mengen von Dingen.» (Man könnte hinzufügen: Die Gegenstände, die wahre Beschreibung und die Entsprechungsbeziehung sind unabhängig von den menschlichen Erkenntnismöglichkeiten zu rekonstruieren). Dem so definierten metaphysischen Realisten ist vor allem vorzuwerfen, daß er seine eigene epistemische Position ignoriert und so tut, als könne er die Welt und das Erkennen der Welt quasi von außen, aus der Perspektive eines Gottes betrachten. Der interne Realist hält mit Kant dagegen, daß wir nur von einer Welt für uns reden können, nicht von den Dingen an sich — auf die Dinge an sich können wir nicht referieren. In diesem Text (der auf einem Kongreß in Österreich vorgetragen wurde) wird der Versuch sehr deutlich, an ‹kontinentale› Traditionen anzuknüpfen und internen Realismus als eine klassisch erkenntnistheoretische Position zu formulieren.

In ‹Warum es keine Fertigwelt gibt› (1982, Text 7) holt Putnam zum großen Schlag gegen die beherrschende metaphysische Auffassung aus, die zum Teil ein Kind seiner eigenen früheren Philosophie ist: den Materialismus. Man hatte die ‹kausale Theorie der Referenz› des frühen Putnam und Kripke (1972) zu einer ‹physikalistischen› oder ‹naturalistischen› Theorie (so die modernen, ‹wissenschaftlichen› Bezeichnungen für Materialismus) ausgebaut und meinte, die Kausalbeziehung zwischen Gegenstand und Ausdruck werde eine Referenzbeziehung auf quasi natürliche Weise auszeichnen. Der Ausdruck referiere auf das, wozu er in einer Kausalbeziehung der richtigen Art steht. Referenz wird auf Kausalität reduziert. Diese ‹Metaphysik innerhalb der Grenzen der bloßen Wissenschaft› versucht, ganz ohne intentionale Begriffe auszukommen.

Putnam will vor allem zeigen, daß der Begriff der Kausalität selbst kein physikalischer Begriff ist — zumindest gelte das für jenen Begriff der ‹erklärenden› Kausalität, den der Materialist benötigt. Kausalität kann also nicht das leisten, was ihr von kausalen Referenztheoretikern aufgebürdet wird. Allgemein verwirft er den metaphysischen Versuch, von den ‹in der Welt da draußen› vorhandenen Gegenständen zu sprechen, von der Welt, die schon fertig vorliegt, bevor wir sie erkennen. Der Materialismus ist der letzte große Entwurf dieser Art, der noch eine nennenswerte Zahl von Anhängern besitzt.

### 4

Der interne Realismus scheint nun genügend vom metaphysischen Realismus abgegrenzt zu sein, und Putnam hat sich ‹kontinentalen› Auffassungen genähert, welche die Frage, «Was kann ich wissen?»[14] mit dem Hinweis auf die Grenzen von Erkenntnis beantworten. Ist Putnam damit aber nicht einem Relativismus verfallen? Gestattet interner Realismus nicht beliebige Wahrheiten, je intern zu einem begrifflichen Schema, das ich mir aussuche? Fällt nicht so alles der Beliebigkeit anheim? Putnam will aus dem Scheitern des metaphysischen Realismus keineswegs diese relativistische Konsequenz ziehen, sondern versucht, auf des Messers Schneide zu wandeln: interner Realismus, aber nicht Relativismus.

Der für eine französische Enzyklopädie geschriebene Text 8 ‹Wozu die Philosophen?› (1986) bietet vor allem eine philosophiehistorische Einordnung von Putnams durch Putnam und eine Rekapitulation der Argumente gegen metaphysischen Realismus. Putnam betont, daß er den Positivismus als eine Spielart des Idealismus ansieht, die im Endeffekt auf Relativismus hinausläuft (seine alten Gegner sind also auch nicht ohne diese Schwierigkeit), was sich an Quines Lehre von der ontologischen Relativität (1969) illustrieren läßt. Er lehnt weiterhin die mißliche Alternative zwischen Reduktionismus (wie dem Positivismus) und metaphysischen Positionen (wie dem Materialismus) ab, sieht aber keinen Grund, daraus zu schließen, es gebe keine besseren und schlechteren Kontexte, innerhalb derer dann Realismus ‹intern› sein kann. Es gibt da vernünftige und weniger vernünftige Entscheidungen.

Außerdem findet sich eine konzise Darstellung seiner berühmten Widerlegung des skeptischen Arguments, daß wir alle uns ständig irren könnten, weil wir etwa ‹Gehirne im Tank› sind. Putnam meint, ein Gehirn im Tank kann nicht auf ein Gehirn im Tank referieren. Bedeutungen sind nicht im Kopf.

Die Frage von verschiedenen gleich guten Beschreibungen, wird in der zentralen Lehre von der ‹begrifflichen Relativität› deutlicher, wie sie vor allem in Text 9 ‹Realismus mit menschlichem Antlitz› (1987/90) entfaltet wird. Es kann durchaus mehrere Arten und Weisen geben, über einen Sachverhalt zu reden, zwischen denen es nichts zu wählen gibt (man erinnere sich an ‹Modelle und Wirklichkeit›). Hierfür werden mehrere Beispiele gegeben, wobei Putnam auf seinen reichen Fundus an wissenschaftlichen Kenntnissen zurückgreift. Er diskutiert die Deutung der Quantenmechanik sowie die Lügnerparadoxie und interpretiert diese als Hinweise auf die Vergeblichkeit der Suche nach der einen wahren Wissenschaft, zu der alles Wissen konvergiert. Es wird immer mehrere Theorien geben, aber das schadet auch gar nichts, solange diese mit verschiedenen Begriffen arbeiten. Solange man sich vor Augen hält, daß Wissenschaft nicht von den Kantischen Dingen *an sich* handelt, sondern von Dingen *für uns*, wird man sich nicht wundern, daß wir verschiedene Theorien haben können, die empirisch äquivalent sind, für uns keinen Unterschied machen — und das heißt eben, es *gibt* keinen relevanten Unterschied zwischen ihnen (was ein metaphysischer Realist bestreiten würde).

Im zweiten Teil des Essays setzt Putnam sich mit Richard Rorty auseinander, einem Philosophen, der aus der analytischen Schule stammt, aber inzwischen das Ende der analytischen Philosophie verkündet (1979) und das ganze Problem der ‹Repräsentation› oder ‹Darstellung› von Welt für fehlgeleitet hält — was bedeutet, er betrachtet den Streit zwischen Antirealisten und Realisten als Scheindebatte. Genau das hält Putnam aber für den Versuch, vom Standpunkt eines Gottes aus zu sagen, es gebe keinen Standpunkt eines Gottes; für den Versuch, metaphysisch zu sagen, es gebe keine Metaphysik. Das Scheitern des Projekts ‹Metaphysik› hat allerdings nach Putnam nicht die große kulturelle Bedeutung, welche Rorty ihr im Anschluß an französische Philosophen zuschreibt.

Auffällig ist, daß sich Putnam in den letzten drei Texten dieses Bandes mit französischer Philosophie beschäftigt. Zumindest wenn er an Relativismus denkt, fallen ihm offensichtlich Namen wie Derrida oder Foucault ein. Die französische Gegenwartsphilosophie gibt noch am ehesten einen Widerpart zur analytischen Philosophie ab, an dem zu reiben sich lohnt.

Ein weiterer Versuch der Näherung stellt Text 10 dar, ‹Irrealismus und Dekonstruktion› (1992). Putnam befaßt sich hier mit der Philosophie seines älteren Kollegen in Harvard, Nelson Goodman, und zugleich mit Jaques Derrida. Als Reaktionen auf Physikalismus und Szientismus sind ihm diese Positionen sympathisch, er meint jedoch, sie schütten das Kind mit dem Bade aus. Zur Bewertung von Goodmans Philosophie diskutiert Putnam die Fragen «Haben wir die Sterne gemacht?» und «Haben wir den Großen Wagen gemacht?», wobei noch einmal das Problem der begrifflichen Relativität verdeutlicht wird. Putnam bleibt Realist, insofern er weiterhin einen wesentlichen Unterschied zwischen Fragen der Konvention (Was gehört zum Großen Wagen?) und Fragen der Fakten in der Welt (Ist das ein Stern?) sieht, der sich wiederum als Unterschied in der jeweils angemessenen Theorie der Bedeutung formulieren läßt. Die Bedeutung von ‹Stern› muß anders rekonstruiert werden als die von ‹Großer Wagen›.

Die Näherung an Derridas ‹Dekonstruktion› fällt sichtlich schwer, aber allein die ernsthafte Auseinandersetzung bezeugt eine für einen führenden analytischen Philosophen noch vor kurzem schwer vorstellbare Wertschätzung. Derrida wird allerdings gleich eines alten Fehlers bezichtigt: Relativisten definieren ‹Wissen› typischerweise mit extrem hohen Standards um dann zu dem Schluß zu kommen, diese könnten nicht erreicht werden. Auch Derridas Kritik am ‹Logozentrismus› hält Putnam nicht nur für verfehlt, sondern sogar für gefährlich; denn sie laufe Gefahr ethisch verwerfliche Handlungsweisen unkritisiert zu lassen. Sein Resumee lautet: «Dekonstruktion ohne Rekonstruktion ist Unverantwortlichkeit.»

In Putnams Denken zeigt sich die allgemeine Annäherung der analytischen Philosophie an ‹kontinentale›, traditionelle Fragestellungen und Autoren. Die führenden Namen in den angloamerikanischen Ländern sind nicht mehr dieselben wie noch von 20 Jahren. Leute wie Putnam und Rorty, Williams, Dennett, Nagel, Kripke und Davidson oder auch Kuhn und Feyerabend haben je auf ihre Weise Annäherung versucht. Der vorliegende Band soll dazu beitragen, daß die Annäherung eine gegenseitige bleibt.

## 5

Wie viele andere angelsächsische Autoren schreibt Putnam (vor allem in seinen früheren Schriften) für ein Publikum, das ‹schon weiß, was gemeint ist›, wenn er einen Namen erwähnt. Die lästige Pflicht, genauere Literaturhinweise zu machen, wurde hier vom Herausgeber nachgeholt. Alles, was nicht von Putnam stammt, ist zur Warnung in eckige Klammern eingeschlossen. Es finden sich dort auch einige Verweise auf andere Stellen bei Putnam, innerhalb und außerhalb dieses Bandes. Erklärende Ausführungen und Erläuterungen zur Übersetzung wird man allerdings vergeblich suchen.

Wie in den jeweiligen Fußnoten mit Sternchen* zu erkennen, geben die Übersetzungen den jeweils letzten autorisierten Text wieder. Das führt teilweise dazu, daß anachronistisch erscheinende, von Putnam später hinzugefügte Verweise auf später erschienene Texte auftauchen. Es wurde versucht, einen deutschen Text so zu schreiben, wie er wohl lauten würde, wenn Putnam zum betreffenden Zeitpunkt als Muttersprachler auf Deutsch geschrieben hätte. Das heißt, weder auf Deutsch so zu schreiben, wie Putnam auf Englisch geschrieben hat, noch das, was er sagen wollte, so zu schreiben, wie es der Übersetzer auf Deutsch geschrieben hätte. Im Register findet sich bei einer Reihe von Einträgen der englische Ausdruck in Klammern. Man kann dort also nachsehen, wie ein deutscher Ausdruck im Englischen lautete.[15]

Hilary Putnam bin ich für Hilfe bei der Bibliographie ab 1992 und vor allem für die Beantwortung zahlreicher Fragen zur Übersetzung

verpflichtet. Der vorliegende Text enthält eine Reihe von daraus resultierenden Korrekturen gegenüber den Originalen. Für vielfältige Unterstützung danke ich besonders Sophia Voulgari, Mark Sainsbury, David Papineau und Stathis Psillos, King's College London; Rom Harré und Peter M. S. Hacker, Oxford; Burghard König und Wolfgang Künne, Hamburg sowie Frank Mathwig, Marburg.

Hamburg, im August 1993

V. C. M.

## Anmerkungen

[1] John Passmore, *Recent Philosophers. A Supplement to A Hundred Years of Philosophy*, Duckworth, London 1985, 97. Passmores sehr lesenswerte Darstellung behandelt die analytische Philosophie zwischen 1965 und 1985.

[2] Bezüge auf Putnams Schriften stets durch den Titel (auf Deutsch, falls Übersetzungen vorhanden sind) und einen Verweis auf die Stelle in der Bibliographie: Jahreszahl der Erstpublikation (1990), Art des Textes (I) und laufende Nummer (13). Auf Literatur anderer Autoren mit Nachname, Jahreszahl. Die relevante Literatur wurde bei den jeweiligen Texten in Anmerkungen nachgeliefert, also bleiben die Verweise in dieser Einleitung spärlich. Echte Sekundärliteratur gibt es ohnehin kaum; man konsultiere aber die Zeitschriftenbände mit Repliken Putnams (*Erkenntnis* 34, 1991 und *Philosophical Topics* 20, 1992) sowie den Band: George Boolos (Hg.): *Meaning and Method: Essays in Honor of Hilary Putnam*, Cambridge University Press, Cambridge 1990. Im Erscheinen sind: Bob Hale/ Peter Clarke (Hg.): *Reading Putnam. Proceedings of the 1990 Gifford Conference at St. Andrews*, Basil Blackwell, Oxford 1993; und auch John Haldane/Crispin Wright (Hg.): *Reality: Representation and Projection*, Oxford University Press, Oxford 1993.

[3] Wie etwa bei *Vernunft, Wahrheit und Geschichte* (1981, I-6) und *Realität und Repräsentation* (1988, I-11).

[4] Der nicht dort abgedruckte Aufsatz von Putnam und Paul Oppenheim über die Einheit der Wissenschaft (1958, II-11) ist 1970 in deutscher Übersetzung erschienen (III-1).

[5] Putnams Schriften zu diesem Thema sind in Anmerkung 21 zu Text 3 aufgelistet.

[6] Siehe etwa ‹The Analytic and the Synthetic› (1962, II-23), ‹How Not to Talk about Meaning. Comments on J. C. C. Smart› (1965, II-38) oder ‹Is Semantics Possible?› (1970, II-57).

[7] Putnam, ‹Introduction: Philosophy of Language and the Rest of Philosophy› (1975, II-77), X.

[8] Eine ausführlichere Fassung der früheren realistischen Theorie der Bedeutung einschließlich einer Auseinandersetzung mit zeitgenössischen Alternativen findet sich in

«Die Bedeutung von ‹Bedeutung›?» (1975, II-86), dessen deutsche Übersetzung als Monographie erschienen ist.

[9] Tarski hatte vor allem ein Adäquatheitskriterium für eine Wahrheitsdefinition aufgestellt: Sie sollte alle Äquivalenzen der Form «‹p› ist wahr dann und nur dann, wenn p» implizieren. Genaueres und Literatur, siehe Text 5, Abschnitt ‹Tarskis Wahrheitstheorie›.

[10] Eine ausführliche Darstellung von Putnams Philosophie zur Bedeutung und Erkenntnis bis 1978 (dem Erscheinen von *Meaning and the Moral Sciences*, 1978, I-5) bietet Wolfgang Stegmüller in Band II, seiner *Hauptströmungen der Gegenwartsphilosophie*, Kröner, Stuttgart, 7. erw. Aufl. 1987, 345–467.

[11] Insbesondere wird die zeitlich frühere und darum bekanntere Präsentation in ‹Realism and Reason› (1977, II-92) ignoriert, weil sich deren Argumente in späteren Texten prägnanter ausgedrückt finden.

[12] Siehe etwa Peirce/Rantala: ‹Realism and Reference›, in: *Synthese* 52 (1982), 439–448 und David Lewis: ‹Putnam's Paradox›, in: *Australasian Journal of Philosophy* 62, 1984, 221–236.

[13] Putnams Auffassung sollte also nicht ohne weiteres mit Konsensustheorien à la Habermas verwechselt werden. Dennoch fällt die Nähe von Habermas zu Diskussionen zwischen Dummett, Putnam und vielen anderen ins Auge. (J. Habermas: ‹Wahrheitstheorien›, in: H. Fahrenbach (Hg.): *Wirklichkeit und Reflexion*, Pfullingen, Neske 1978.)

[14] Kant in der *Logik* (Jäsche), A 26 bringt die Philosophie auf die Fragen: «Was kann ich wissen? Was soll ich tun? Was darf ich hoffen? Was ist der Mensch?» (*Werke*, Bd. VI, hg. v. W. Weischedel, Suhrkamp, Frankfurt/Main 1968.)

[15] Eine gewisse Kontinuität zu den Übersetzungen von W. Spohn und J. Schulte ist dabei gewahrt. Allerdings wurde ‹reference› hier mit dem üblichen Fachausdruck ‹Referenz› übersetzt, während Schulte von ‹Bezug› spricht. ‹Referenz› scheint sich im Deutschen fest etabliert zu haben und es gibt keinen zwingenden Grund, hier sprachpädagogisch tätig zu werden, auch wenn ‹Bezug› sicher glatter klingt. Der Ausdruck ‹Bezug› hat daneben den Nachteil, die *Handlung* des Sich-Beziehens zu betonen — was dann manche Referenztheorien merkwürdiger aussehen läßt, als sie sind. Es steht zu hoffen, daß Putnam-Lektüre dabei hilft, folgenden Fehlschluß zu vermeiden: «Wäre ‹reference› ein Terminus technicus, könnte ihn jeder nach Belieben und gemäß dem zu erwartenden Nutzen definieren, ohne daß eine inhaltliche Auseinandersetzung über den Begriff möglich wäre. Es gäbe also gar keine philosophische Problematik der Bezugnahme!» (J. Schulte: ‹Peter Frederick Strawson›, in: A. Hügli/P. Lübke (Hg.): *Philosophie im 20. Jahrhundert*, Band 2, *Wissenschaftstheorie und Analytische Philosophie*, Rowohlt, Reinbek 1993, S. 446, Anm. 1.)